\documentclass[12pt]{article}
\usepackage{amsmath,amssymb}
\usepackage{graphicx}
\begin{document}

\begin{center}

{\Large\bf Asteroseismology of Vibration Powered Neutron Stars}

 \vspace*{1.0cm}

{\sf 
Sergey  Bastrukov$^{1,2}$, Renxin Xu$^{1}$, Junwei Yu$^{1}$,\\[0.1cm]

Irina Molodtsova$^{2}$ and Hsiang-Kuang Chang$^3$

\vspace*{1.0cm}

$^1${State Key Laboratory of Nuclear Physics and Technology,\\
 School of Physics, Peking University, 100871 Beijing, China\\
 $^2$Joint Institute for Nuclear Research, 141980 Dubna, Russia\\
 $^3$ Institute of Astronomy and Department of Physics,\\
National Tsing Hua University, 
Hsinchu 30013, Taiwan}

}
\end{center}

 \vspace*{7.5cm}

\sf

\noindent
{\large In the book "Astrophysics" Chapter 13, p.287-308. \\
\noindent
Edited by Ibrahim Kucuk, ISBN 978-953-51-0473-5, InTech, March 3, 2012.}

\newpage

\section{Introduction}
 There is a general recognition today that basic features of asteroseismology of 
  non-convective final stage (FS) stars, such as white dwarfs, neutron stars and 
  and strange quark star, can be 
  properly understood working from the model of vibrating solid star, rather than 
  the liquid star model lying at the base of asteroseismology 
  of convective main-sequence (MS) stars. In accord with this, most of 
  current investigations of an FS-star vibrations rests on  
  principles of solid-mechanical theory of continuous media, contrary to the study of  
  the MS-star vibrations  which are treated in terms of fluid-mechanical 
  theory.
  This means that super dense matter of FS-stars (whose gravitational pressure 
  is counterbalanced by degeneracy pressure of constituting Fermi-matter), 
  possesses elasticity and viscosity generic to solid state of condensed matter, 
  whereas a fairly dilute matter of the MS (whose internal pressure of self-gravity is 
  opposed by radiative pressure) possesses property of fluidity which is generic 
  to the liquid state of a highly conducting condensed matter. This feature 
  of the MS-star matter plays crucial role in generation of their 
  magnetic fields in the dynamo processes involving macroscopic flows which are 
  supported by energy of nuclear reactions in the central, reactive zone, of these
  stars. In the meantime, in the FS stars, like white dwarfs and neutron 
  stars, there are no nuclear energy sources to support convection. The prevailing today view, therefore, is that a highly stable to spontaneous decay dipolar magnetic fields of neutron stars are fossil. The extremely large intensity of magnetic fields 
  of degenerate solid stars is attributed to amplification of fossil magnetic field 
  in the magnetic-flux-conserving process of core-collapse supernova.
  
  Even still before discovery of neutron stars, it has been realized that, for 
  absence of nuclear sources of energy, the radiative activity of these of compact 
  objects should be powered by energy of stored in either rotation or vibrations
  and that the key role in maintaining the neutron star radiation should play an 
  ultra strong magnetic field. As is commonly know today, the neutron star capability 
  of accommodating such a field is central to
 understanding pulsating character of magneto-dipole radiation of radio pulsars
 whose radiative power is provided, as is commonly believed, by the energy of
 rigid-body rotation. The discovery of soft gamma-ray repeaters and
 their identification with magnetars \cite{DT-92}  -- quaking neutron stars endowed
 with ultra strong magnetic fields experiencing decay -- has stimulated enhanced
 interest in the study of models of quake-induced
 magneto-mechanical seismic vibrations of neutron star and resultant
 electromagnetic radiation.  Of particular interest in this study are
 torsional magneto-mechanical vibrations about axis of magnetic dipole moment of
 the star driven by forces of magnetic-field-dependent stresses in a perfectly conducting
 matter and in a permanently magnetized non-conducting matter \cite{B-WS}.  Most, if not all,  reported up to now computations of frequency spectra of poloidal
 and toroidal Alfv\'en vibration modes in pulsars and magnetars rest
 on tacitly adopted assumption about constant-in-time magnetic field in
 which a perfectly conducting neutron star matter undergoes Lorentz-force-driven
 oscillations \cite{Car-86,BP-97,B-99b,Lee-07,Lee-08,
 B-07a,B-07b,B-09a,B-09b,B-10c} (see, also, references therein). A special place in the study of the above Alfv\'en
 modes of pure shear magneto-mechanical seismic vibrations ($a$-modes) occupies a homogeneous model of a solid star with the uniform density
 $\rho$ and frozen-in poloidal static magnetic field of both homogeneous and
 inhomogeneous internal and dipolar external configuration, and we start section 2 with a brief outline of this model.
 In section 3, a mathematical background
  of quaking neutron star model is briefly outlined with emphasis on the loss of
  vibration energy caused by depletion of internal magnetic field pressure  and
  resulting vibration-energy powered magneto-dipole radiation.
  The decreasing of magnetic field pressure in the star is presumed to be caused by
  coupling between vibrating star and outgoing material which is expelled by quake,
  but mechanisms of star-envelope interaction resulting in the decay of
  magnetic field, during the time of vibrational relaxation, are not considered.
   It is shown that physically meaningful inferences regarding radiative activity
  of quaking neutron star can be made even when detailed mechanisms of depletion
  of magnetic field pressure in the process of vibrations triggered by quakes are not
  exactly known. This statement is demonstrated by a set of representative examples  of magnetic field decay. The basic results of the model of vibration powered
  neutron star are summarized in section 4 with emphasis of its relevance to 
  astrophysics of magnetars. 
 
 \section{Lorentz-force-driven torsion vibrations of a neutron star}
 
 To gain better understanding of the basic physics behind the interconnection 
 between seismic and radiative activity of quaking neutron star, we start with a brief 
 outline a fiducial model of a solid star with frozen-in homogeneous internal 
magnetic field 
  \begin{eqnarray}
  \label{e1.1}
 && {\bf B}({\bf r})=B\,{\bf b}({\bf r}),\quad B={\rm constant},\\
 \nonumber
 && {\bf b}({\bf r})=[b_r=\cos\theta,
 b_\theta=-\sin\theta, b_\phi=0]
  \end{eqnarray}
   and dipolar external magnetic field,  as is shown in Fig.1. In the last  
   equation, $B$ is the field intensity [in Gauss] and
   ${\bf b}({\bf r})$ stands for the dimensionless vector-function of spatial
   distribution of the field.  
     \begin{figure}
\centering
 \includegraphics[scale=0.3]{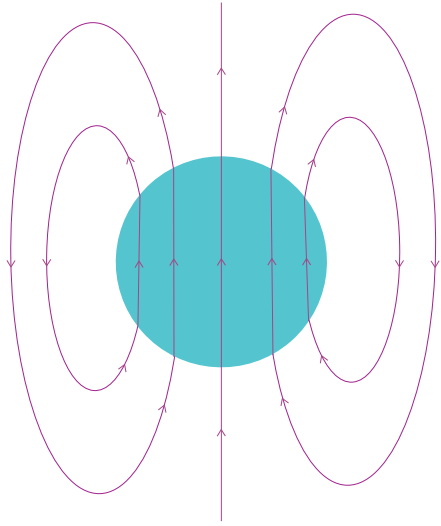}
   \caption{The lines of magnetic field in fiducial model of neutron star 
	   with dipolar external and homogeneous internal  magnetic field.} 
\end{figure}  
 
 The above form of ${\bf B}({\bf r})$ has
   been
   utilized in recent works\cite{B-09a,B-09b,B-10c} in which the
   discrete spectra of frequencies of node-free
   torsional Alfv\'en oscillations has been computed in analytic form on the basis of
   equations of linear magneto-solid mechanics
 \begin{eqnarray}
  \label{e1.1a}
 && \rho{\ddot {\bf u}}=\frac{1}{c}
 [\delta {\bf j}\times {\bf B}],\quad \nabla\cdot {\bf u}=0,\\
  \label{e1.1b}
 && \delta {\bf j}=\frac{c}{4\pi}[\nabla\times \delta {\bf B}],\,\,
 \delta {\bf B}=\nabla\times [{\bf u}
  \times {\bf B}],\,\,\nabla\cdot {\bf B}=0.
  \end{eqnarray}
   Here ${\bf u}={\bf u}({\bf r},t)$ is the field of material displacement, the
   fundamental dynamical variable of solid-mechanical theory of elastically deformable
   (non-flowing) material continua.  These equations describe  Lorentz-force-driven
    non-compressional vibrations of a perfectly conducting elastic matter of a non-
    convective solid star
    with Amp\'ere form of  fluctuating current density $\delta {\bf j}$. Equation for
    $\delta
    {\bf B}$ describing coupling between fluctuating field of
    material displacements ${\bf u}$
    and background magnetic field ${\bf B}$ pervading stellar material is the
    mathematical form of Alfv\'en
    theorem about frozen-in lines of magnetic field in perfectly conducting
    matter. The adopted for Alfv\'en vibrations of non-convective solid stars
    terminology [which is
    not new of course, see for instance\cite{M-84}, namely, equations of
    magneto-solid mechanics  and/or  solid-magnetics is used as a solid-mechanical
    counterpart  of well-known terms like equations of magneto-fluid mechanics,
    magnetohydrodynamics (MHD) and
    hydromagnetics  $\rho \delta {\dot {\bf v}}=(1/c)
  [\delta {\bf j}\times {\bf B}],\quad \delta {\bf j}=(c/4\pi)[\nabla\times \delta {\bf
  B}],\quad \delta {\dot {\bf B}}=\nabla\times [\delta {\bf v}
  \times {\bf B}]$, describing flowing magento-active plasma in terms of the velocity
  of fluctuating flow $\delta {\bf v}={\dot {\bf u}}$  and fluctuating magnetic field
  ${\delta {\bf B}}$.
   The MHD approach is normally utilized in astrophysics of convective main-sequence (MS) liquid stars such, for instance,  as
   rapidly oscillating Ap (roAp) stars, chemically peculiar magnetic
stars exhibiting high-frequency oscillations that have been and still remain
the subject of extensive investigation\cite{LW-58,RR-03}.

        Inserting (\ref{e1.1a}) in (\ref{e1.1b})
  the former equation of solid-magnetics takes the form
    \begin{eqnarray}
  \label{e1.2}
 && \rho\, {\ddot {\bf u}}({\bf r},t)=\frac{1}{4\pi}
 [\nabla\times[\nabla\times [{\bf u}({\bf r},t)\times {\bf B}({\bf r})]]]\times {\bf B}
 ({\bf r}).
  \end{eqnarray}
   The analogy between perfectly conducting medium pervaded 
  by magnetic field (magneto-active plasma) 
  and elastic solid, regarded as a material continuum, is 
  strengthened by the following tensor representation of the last equation
   $\rho\,{\ddot u}_{i}=\nabla_k \delta M_{ik}$, where $\delta M_{ik}=(1/4\pi)
   [B_i\delta 
  B_k+B_k\delta B_i-B_j\delta B_j\delta_{ik}]$ is the Maxwellian tensor of  
 magnetic field stresses with  $\delta B_i = \nabla_k[u_i B_k - u_k B_i]$. 
 This form is identical in appearance to canonical equation of solid-mechanics 
 $\rho\,{\ddot u}_{i}=\nabla_k \sigma_{ik}$, where
 $\sigma_{ik}=2\mu\,u_{ik}+[\kappa-(2/3)\mu]\,u_{jj}\delta_{ik}$ is the
 Hookean tensor of mechanical stresses and $u_{ik}=(1/2) [\nabla_{i}\,u_k+\nabla_k\,u_i]$ is the 
  tensor of shear deformations in an isotropic elastic continuous matter with shear
  modulus $\mu$ and bulk modulus $\kappa$ (having physical dimension of 
  pressure). 
  The most prominent manifest of physical similarity between 
  these two material continua is their capability of transmitting non-compressional 
  perturbation by transverse waves.  Unlike incompressible liquid, 
  an elastic solid can respond to impulsive non-compressional load by 
  transverse waves of shear material displacements traveling with the speed 
  $c_t=\sqrt{\mu/\rho}$. The unique feature of an incompressible perfectly 
  conducting and magnetized continuous matter (in liquid or solid aggregated state) 
  is the capability of transmitting perturbation by transverse magneto-mechanical,  
  Alfv\'en, wave in which mechanical displacements of material and fluctuations 
  of magnetic field undergo coupled oscillations traveling with
  the speed $v_A=B/{\sqrt{4\pi\rho}}=\sqrt{2P_B/\rho}$ (where 
  $P_B=B^2/8\pi$ is magnetic field pressure) along magnetic axis.  It is stated, 
  therefore, that magnetic field  
  pervading perfectly conducting medium imparts to it a supplementary portion of 
  solid-mechanical elasticity \cite{Ch-61,AF-63}. This suggests that hydromagnetic 
  Alfv\'en vibrations of a spherical mass of a perfectly conducting matter with frozen-in 
  magnetic field can be specified in a manner of eigenstates of elastic vibrations of a 
  solid sphere.  As for the general asteroseismology of compact objects is concerned, the above
    equations  seems to be appropriate not only for neutron stars but also white
    dwarfs\cite{M-10} and quark stars. The superdense material of these
   latter yet hypothetical compact stars is too
   expected to be in solid state\cite{Xu-03,Xu-09}.

    Equation (\ref{e1.2}) serves as a basis of our further analysis. The studied in
  above works regime of node-free torsion vibrations under the action of Lorentz restoring force is of some interest in that the rate
  of differentially rotational material displacements
   \begin{eqnarray}
   \label{e1.3}
 && {\dot {\bf u}}({\bf r},t)=
  [\mbox{\boldmath $\omega$}({\bf r},t)\times {\bf r}],\quad
  \mbox{\boldmath $\omega$}({\bf r},t)=[\nabla\chi({\bf r})]\,{\dot\alpha}(t), \\
   \label{e1.4}
  && \nabla^2\chi({\bf
  r})=0,\quad \chi({\bf r})=f_\ell\,P_\ell(\cos\theta), \quad f_\ell(r)=A_\ell\,
  r^{\ell}
 \end{eqnarray}
  has one and the same form as in torsion elastic mode of node-free
  vibrations under the action of Hooke's force of mechanical
  shear stresses.
  
   \begin{figure}
\centering
 \includegraphics[scale=0.5]{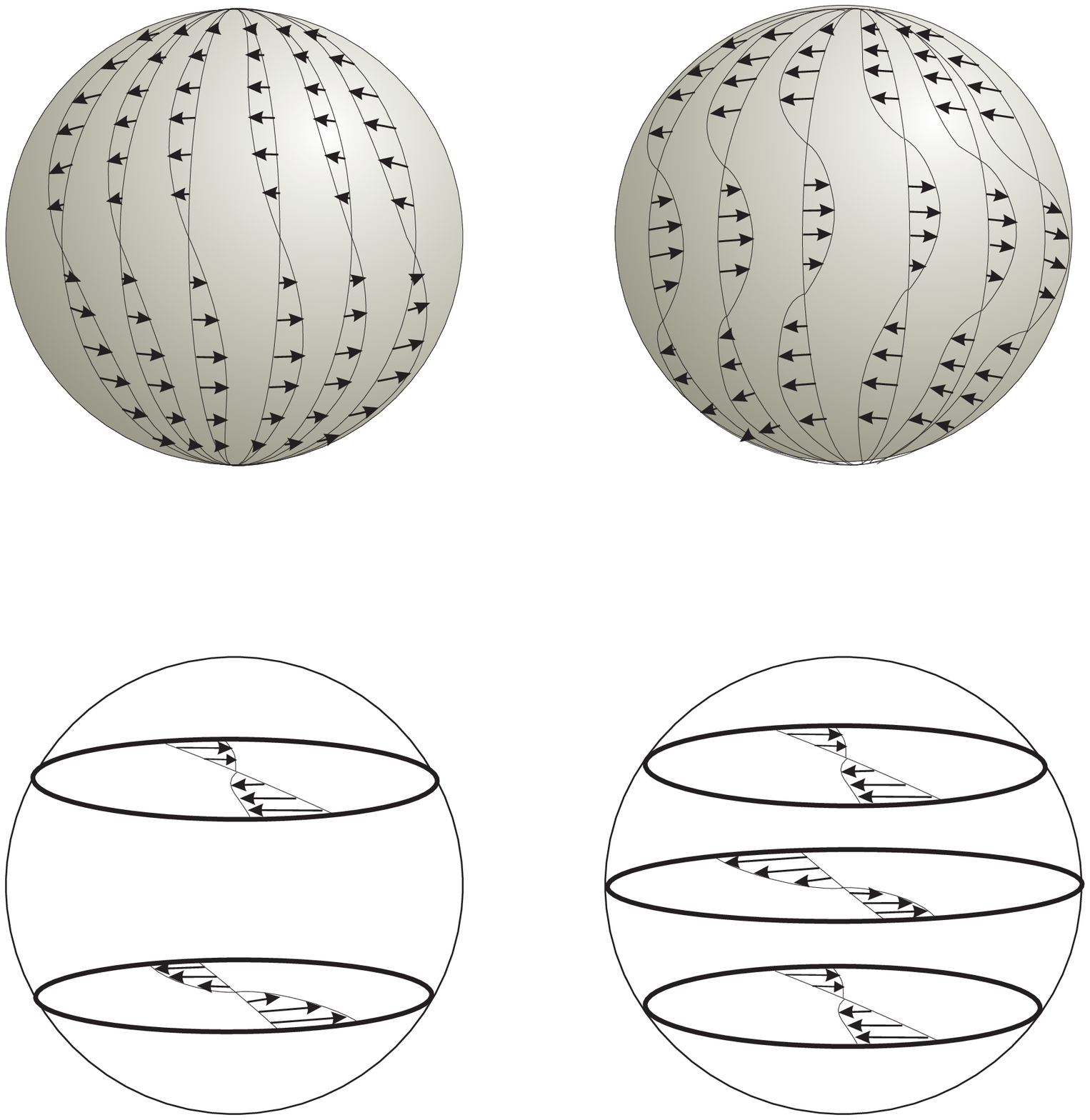}
   \caption{The fields of material displacements in a neutron star undergoing torsional 
   quadrupole (left) and octupole (right) node-free vibrations about magnetic axis.} 
\end{figure}

  Hereafter $P_\ell(\cos\theta)$ stands for
  Legendre polynomial of degree $\ell$ specifying the overtone of toroidal
  $a$-mode. Fig.2 shows quadrupole and octupole overtones of such 
  vibrations. The
  time-dependent amplitude $\alpha(t)$ describes temporal evolution of above
  vibrations; the governing equation for $\alpha(t)$ is obtained form equation
  (\ref{e1.2}).  The prime purpose of above works
  was to get some insight into difference between spectra of discrete frequencies
  of toroidal $a$-modes in neutron star models having one and the same
  mass  $M$ and radius $R$, but different shapes of constant-in-time poloidal
  magnetic fields. By use of the energy method, it was found that each specific
  form of spatial configuration
  of static magnetic field  about axis of which the neutron star matter undergoes
  nodeless torsional oscillations is uniquely reflected in the discrete frequency spectra
  by form of dependence of frequency upon overtone $\ell$ of nodeless vibration.
   It worth noting that first computation
  of discrete spectra of frequencies of toroidal Alfv\'en stellar vibrations in the standing
  wave-regime, has been reported by  Chandrasekhar\cite{Chandra-56}.  The
  extensive review of other earlier computations of discrete frequency spectra of $a$-modes, $\omega_\ell=\omega_A\,s_\ell=B\kappa_\ell$, is given
 in well-known review of Ledoux and Walraven\cite{LW-58}.

\begin{figure}
\centering
 \includegraphics[scale=0.6]{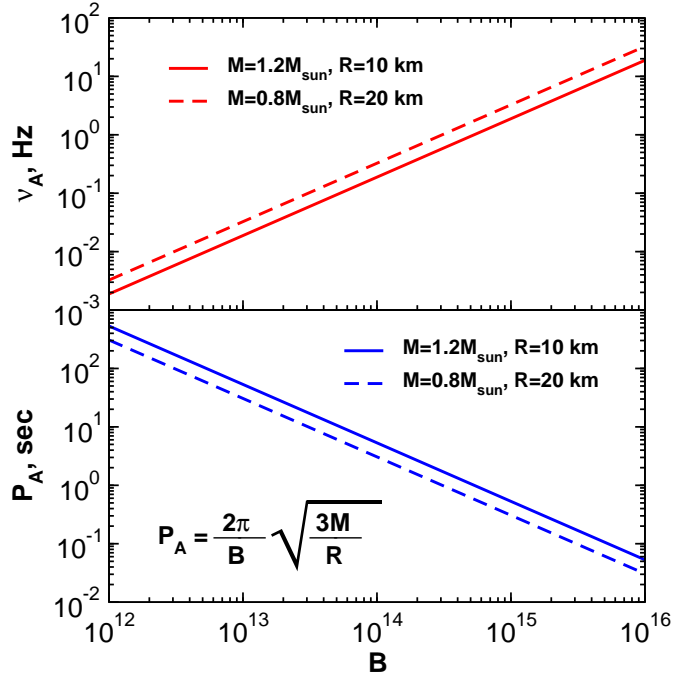}
   \caption{
  The basic frequency and period of global Alfv\'en oscillations, equations
  (\ref{e1.6}), as functions of magnetic field intensity in the neutron star models with
  indicated mass and radius.}
  \label{fig:fr-period}
\end{figure}
  The assumption about constant in time undisturbed magnetic field
  means that the internal magnetic field pressure, $P_B$, the velocity $v_A$ of
  Alfv\'en wave in the star bulk
   \begin{eqnarray}
  \label{e1.5}
   P_B=\frac{B^2}{8\pi}, \quad v_A=\sqrt{\frac{2P_B}{\rho}}
  \end{eqnarray}
  and, hence, the frequency $\nu_A=\omega_A/2\pi$
  (where $\omega_A=v_A/R$) and the period
  $P_A=\nu_A^{-1}$ of global Alfv\'en oscillations
  \begin{eqnarray}
  \label{e1.6}
 \nu_A=\frac{B}{2\pi}\sqrt{\frac{R}{3M}},\quad
 P_A=\frac{2\pi}{B}\sqrt{\frac{3M}{R}}
\end{eqnarray}
 remain constant in the process of vibrations whose amplitude $\alpha(t)$
 subjects to standard equation of undamped harmonic oscillator\cite{B-09a,B-09b}. The allow for viscosity of stellar material leads to
 exponential damping of amplitude, but the frequency $\nu_A$
 and, hence, the period $P_A$ preserve one and
 the same values as in the case of non-viscous vibrations\cite{B-10c}. In
 Fig.3 these latter quantities are plotted as functions of intensity $B$ of undisturbed
 poloidal magnetic field in the neutron star models with indicated mass  $M$ and
 radius $R$.  The practical usefulness of chosen logarithmic scale in this figure is that
 it shows absolute vales of $\nu_A$ and $P_A$ for global vibrations of typical in mass and radius neutron stars.

  In what follows,  we relax
  the assumption about  constant-in-time magnetic field and examine the impact of its decay
  on the  vibration energy and period. A brief analysis  of such a case
  has been given in the context of
  magnetic white dwarfs\cite{B-10a}. In this paper we present a highly extensive consideration
  of this problem in the context of neutron stars with emphasis
  on its relevance to the post-quake radiation of magnetars.
  
 \section{Vibration-energy powered
   magneto-dipole emission of neutron star}

 In the following we consider a model of a neutron star 
 with time-dependent intensity of homogeneous poloidal
 magnetic field which can be conveniently represented in the form
  \begin{eqnarray}
  \label{e2.1}
 && {\bf B}({\bf r},t)=B(t)\,{\bf b}({\bf r}),\quad [b_r=\cos\theta,
 b_\theta=-\sin\theta, b_\phi=0].
  \end{eqnarray}
       On account of this
  the equation of solid-magnetics, (\ref{e1.2}), takes the form
  \begin{eqnarray}
 \label{e2.2}
 && \rho\, {\ddot {\bf u}}({\bf r},t)=\frac{B^2(t)}{4\pi}\,
 [\nabla\times[\nabla\times [{\bf u}({\bf r},t)\times {\bf b}({\bf r},t)]]]\times {\bf b}
 ({\bf r}).
  \end{eqnarray}
 Inserting here the following separable form of fluctuating material
 displacements
\begin{eqnarray}
  \label{e2.3}
 && {\bf u}({\bf r},t)={\bf a}({\bf r})\,\alpha(t)
 \end{eqnarray}
 we obtain
  \begin{eqnarray}
  \label{e2.4}
 && \{\rho\, {\bf a}({\bf r})\}\,{\ddot {\alpha}}(t)=2P_B(t)\,
  \times\{[\nabla\times[\nabla\times [{\bf a}({\bf r})\times {\bf b}({\bf r})]]]\times {\bf b}
 ({\bf r})\}\,{\alpha}(t).
  \end{eqnarray}
 Scalar product of (\ref{e2.4}) with the time-independent field of instantaneous
 displacements ${\bf a}({\bf r})$ followed by integration over the star volume
 leads to equation for amplitude $\alpha(t)$ having the form of equation of oscillator
 with depending on time spring constant
 \begin{eqnarray}
  \label{e2.5}
 && {\cal M}{\ddot \alpha}(t)+{\cal K}(t)\alpha(t)=0,\\
  \label{e2.5a}
 &&  {\cal M}=\rho\,m_\ell,\quad {\cal K}(t)=2P_B(t)\, k_\ell,\\
  \label{e2.6}&&  m_\ell=\int \,{\bf a}({\bf r})\cdot {\bf a}({\bf r})\,d{\cal V},\,
  {\bf a}=A_t\nabla\times [{\bf r}\,r^\ell\,P_\ell(\cos\theta)],\\
  \label{e2.7}
 &&
 k_\ell=\int
 {\bf a}({\bf r})\cdot [{\bf b}({\bf r})\times [\nabla\times[\nabla\times [{\bf a}({\bf
 r})\times {\bf b}({\bf r})]]]]d{\cal V}.
 \end{eqnarray}
 The solution of equation of non-isochronal  (non-uniform in
 duration) and non-stationary vibrations with time-dependent frequency [${\ddot
 \alpha}(t)+\omega^2(t)\alpha(t)=0$ where $\omega^2(t)={\cal K}(t)/{\cal M}$] is
 non-trivial and fairly formidable task\cite{VV-77}.
 But solution of such an equation, however, is not a prime purpose of this work.
 The main subject is the impact of depletion of magnetic-field-pressure on the total
 energy of Alfv\'en vibrations $ E_A=(1/2)[{\cal M}{\dot \alpha}^2+{\cal
 K}\alpha^2]$  and the discrete spectrum of frequency of the toroidal $a$-mode
 \begin{eqnarray}
 \label{e2.9}
 &&\omega_\ell^2(t)=\omega_A^2(t)\,s_\ell^2,\,
 \omega_A^2(t)=\frac{v_A^2(t)}{R^2},\, s_\ell^2=\frac{k_\ell}{m_\ell}\,R^2,\\
 \label{e2.10}
 && \omega_\ell^2(t)=B^2(t)\kappa_\ell^2,\quad \kappa_\ell^2=\frac{s_\ell^2}
 {4\pi\rho R^2},\quad s_\ell^2=\left[(\ell^2-1)\frac{2\ell+3}{2\ell-1}\right],\quad
 \ell\geq 2.
  \end{eqnarray}
  It is to be stated clearly from the onset that it is not our goal here to speculate
  about possible mechanisms of neutron star demagnetization and advocate
  conceivable laws of magnetic field decay. The main purpose is  to
  gain some insight into the effect of arbitrary law of magnetic field decay in quaking
  neutron star on period of Lorentz-force-driven torsional seismic 
  vibrations (whose quadrupole and octupole overtones are pictured 
  in Fig.2) and radiative activity of the star brought about by such vibrations.
 In the reminder of the paper we focus on the case of torsional Alfv\'en vibrations
  in quadrupole ($\ell=2$) overtone. In so doing we omit index $\ell$ putting
  $\omega(t)=\omega_{\ell=2}(t)$.

\subsection{Magnetic-field-decay induced loss of vibration energy}

   The total energy stored in quake-induced Alfv\'en seismic vibrations of the star is
   given by
    \begin{eqnarray}
 \label{e3.1}
  && E_A(t)=\frac{{\cal M}{\dot \alpha}^2(t)}{2}+\frac{{\cal K}(B(t))\alpha^2(t)}
  {2},\\
  && {\cal K}(B(t))=\omega^2(B(t)){\cal M}.
  \end{eqnarray}
  Perhaps most striking consequence of the magnetic-field-pressure depletion
  during the post-quake vibrational relaxation of neutron star is that it leads
  to the loss of vibration energy at a rate proportional to the rate of magnetic field
  decay
  \begin{eqnarray}
 \nonumber
&& \frac{dE_A(t)}{dt}=\dot\alpha(t)[{\cal M}{\ddot \alpha}(t)+{\cal
 K}(B(t))\alpha(t)]+
  \frac{\alpha^2(t)}{2}\frac{d{\cal K}(B(t))}{dt}\\
   \label{e3.1a}
&&= \frac{{\cal M}\alpha^2(t)}{2}\frac{d\omega^2(B)}{dB}\frac{dB(t)}{dt}={\cal
  M}\kappa^2\alpha^2(t)B(t)\frac{dB(t)}{dt}.
  \end{eqnarray}
    In the model under consideration, a fairly rapid decay of magnetic field during the
  time of post-quake vibrational relaxation of the star is though of as caused, to a
  large extent, by coupling of the vibrating star with material expelled by quake.
  In other words, escaping material removing a part of magnetic flux density from
  the star is considered to be a most plausible reason of depletion of internal
  magnetic field pressure in the star.  It seems quite likely that, contrary to viscous
  dissipation, the loss of vibration energy due to decay of magnetic field must be
  accompanied by coherent (non-thermal) electromagnetic radiation. Adhering to this
  supposition in the next section special consideration is given to the conversion of the energy
  of  Lorentz-force-driven
   seismic vibrations into the energy of magneto-dipole emission whose flux
   oscillates with frequency of torsional Alfv\'en magneto-mechanical vibrations of
   the final stage solid stars, like magnetic white dwarf and
   neutron stars.
    Fig.4 replicates seismic torsional Alfv\'en vibrations of the
  star which are accompanied by oscillations of lines of dipolar magnetic field
  defining the beam direction of outburst X-ray emission.

\begin{figure}
 \centering
 \includegraphics[scale=0.4]{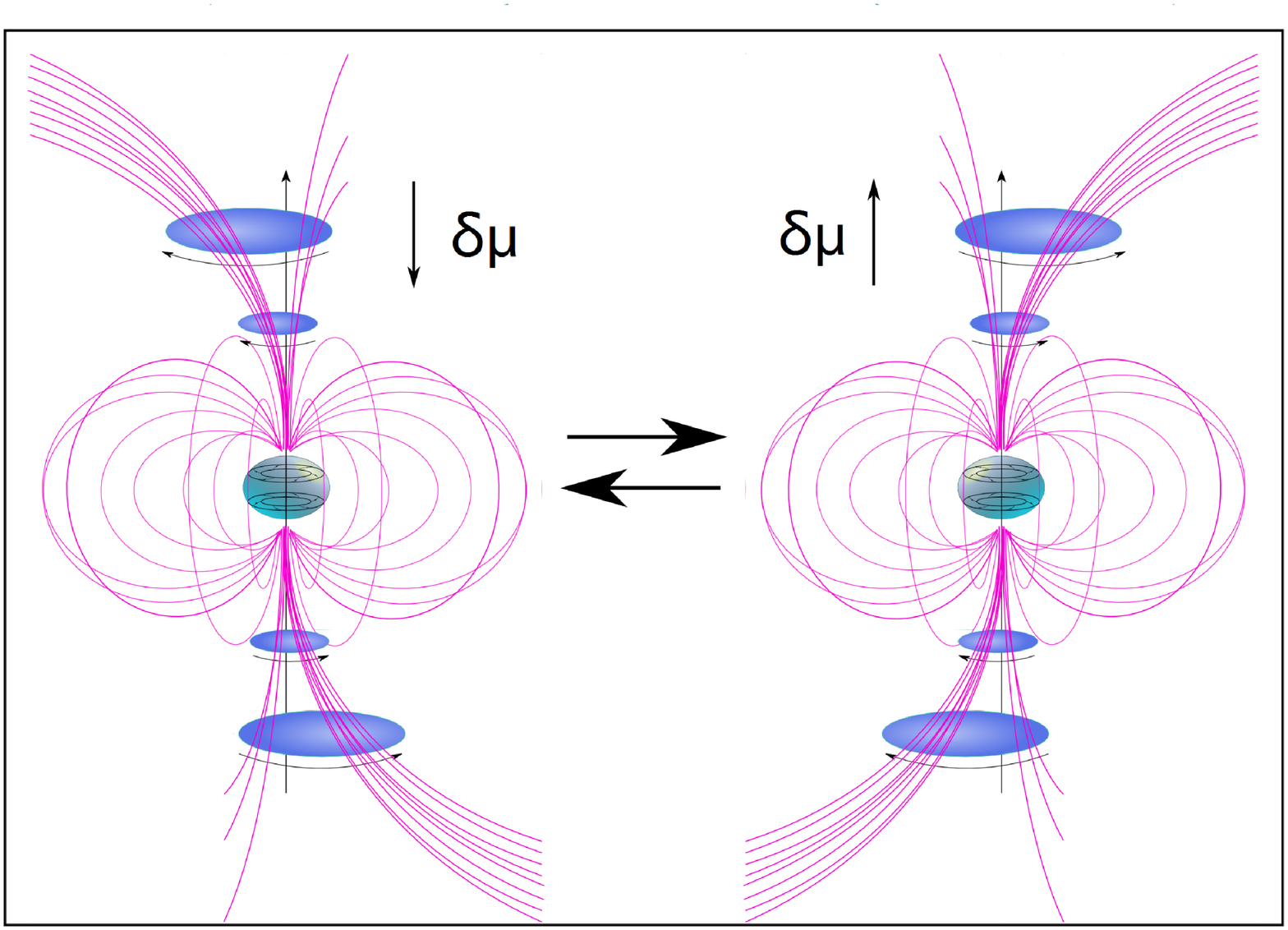}
  \caption{
  Schematic view of quadrupole overtone of seismic torsional Alfv\'en vibrations of
  neutron star
  with homogeneous internal and dipolar external field whose lines, defining
  direction of outburst beam, oscillate with the frequency of this seismic $a$-mode.}
  \label{fig:fr-period}
\end{figure}

 \subsection{Conversion of vibration energy into power of magneto-dipole
   radiation}

    The point of departure in the study of vibration-energy powered
    magneto-dipole emission of the star (whose radiation power, ${\cal P}$, is given by Larmor's
    formula) is the equation
 \begin{eqnarray}
 \label{e3.4}
  && \frac{dE_A(t)}{dt}=-{\cal P}(t),\quad {\cal P}(t)=\frac{2}{3c^3}\delta {\ddot {\mbox{\boldmath
  $\mu$}}}^2(t).
  \end{eqnarray}
   Consider a model of quaking neutron star whose torsional magneto-mechanical
  oscillations are accompanied by fluctuations of total magnetic moment preserving
  its initial (in seismically quiescent state) direction: $\mbox{\boldmath
  $\mu$}=\mu\,{\bf n}={\rm constant}$.
  The total magnetic dipole moment should execute oscillations with frequency
  $\omega(t)$ equal to that for magneto-mechanical vibrations of  stellar matter which
  are described by equation for $\alpha(t)$.
  This means that $\delta {\mbox{\boldmath $\mu$}}(t)$ and  $\alpha(t)$
  must obey equations of similar form, namely
  \begin{eqnarray}
   \label{e3.5}
  && \delta {\ddot {\mbox{\boldmath $\mu$}}}(t)+\omega^2(t)
  \delta {\mbox{\boldmath $\mu$}}(t)=0,\\
   \label{e3.6}
  && {\ddot \alpha}(t)+\omega^2(t){\alpha}(t)=0,\quad \omega^2(t)=B^2(t)
  {\kappa}^2.
  \end{eqnarray}
  It is easy to see that equations (\ref{e3.5}) and  (\ref{e3.6}) can be
  reconciled if
  \begin{eqnarray}
   \label{e3.7}
  \delta \mbox{\boldmath $\mu$}(t)=\mbox{\boldmath $\mu$}\,\alpha(t).
  \end{eqnarray}
  Then, from (\ref{e3.5}), it follows
  $\delta {\ddot {\mbox{\boldmath $\mu$}}}=-\omega^2\mbox{\boldmath $
  \mu$}\alpha$.
  Given this and equating
 \begin{eqnarray}
 \label{e3.8}
  && \frac{dE_A(t)}{dt}={\cal M}\kappa^2\alpha^2(t)B(t)\frac{dB(t)}{dt}
  \end{eqnarray}
  with
  \begin{eqnarray}
   \label{e3.9}
  && -{\cal P}=-\frac{2}{3c^3}\mu^2\,\kappa^4 B^4(t){\alpha}^2(t)
  \end{eqnarray}
  we arrive at the equation of  time evolution of magnetic field
   \begin{eqnarray}
  \label{e3.10}
  && \frac{dB(t)}{dt}=-\gamma\,B^3(t),\quad
  \gamma=\frac{2\mu^2\kappa^2}{3{\cal M}c^3}={\rm
  constant}
  \end{eqnarray}
  which yields the following law of its decay
  \begin{eqnarray}
   \label{e3.11}
  B(t)=\frac{B(0)}{\sqrt{1+t/\tau}},\quad \tau^{-1}=2\gamma B^2(0).
  \end{eqnarray}
  The lifetime of magnetic field $\tau$  is regarded as a parameter
  whose value is established from below given relations
  between the period $P$ and its time derivative
  ${\dot P}$ which are  taken from
  observations. Knowing from observations
 $P(t)$ and ${\dot P}(t)$ and estimating $\tau$ one can get information
 about the magnitude  of total
 magnetic moment and the strength of undisturbed magnetic field.

 It is worth noting that
 in the model of vibration-energy powered
  magneto-dipole emission under consideration, the equation
  of magnetic field evolution  is obtained in similar fashion
  as equation for the angular velocity $\Omega$ does in
  the standard model of rotation-energy powered emission of neutron star.
  As is shown in the next section, the substantial physical difference between
  models of rotation-energy and vibration-energy powered pulsating emission of neutron stars is that in the model of quaking neutron star vibrating in toroidal
  $a$-mode, the elongation of period of pulses is attributed to magnetic field decay,
  whereas in canonical Pacini-Gold model of radio-pulsar the lengthening of period of
  pulses is ascribed to the slow down of the neutron star 
  rotation\cite{MT-77,LK-04,BK-02}.

\subsection{Lengthening of vibration period}

 The immediate consequence of above line of argument is the
  magnetic-field-decay induced lengthening of vibration period
   \begin{eqnarray}
 \label{e3.12}
&&  B(t)=\frac{B(0)}{\sqrt{1+t/\tau}}\,\,\to\,\,P(t)=\frac{C}{B(t)},\,\, C=\frac{2\pi}{\kappa},\\
 && P(t)=P(0)\,[1+(t/\tau)]^{1/2},\,\,  P(0)=\frac{C}{B(0)},\\
 \label{e3.13}
 && {\dot P}(t)=\frac{1}{2\tau}\frac{P(0)}{[1+(t/\tau)]^{1/2}}.
 \label{e3.14}
 \end{eqnarray}
  It follows that lifetime $\tau$ is determined by
 \begin{eqnarray}
 \label{e3.15}
 && {P(t)}{\dot P}(t)=\frac{P^2(0)}{2\tau}={\rm constant}.
 \end{eqnarray}
  This inference of the
  model under consideration is demonstrated in Fig.5.
  \begin{figure}
 \centering
 \includegraphics[scale=0.28]{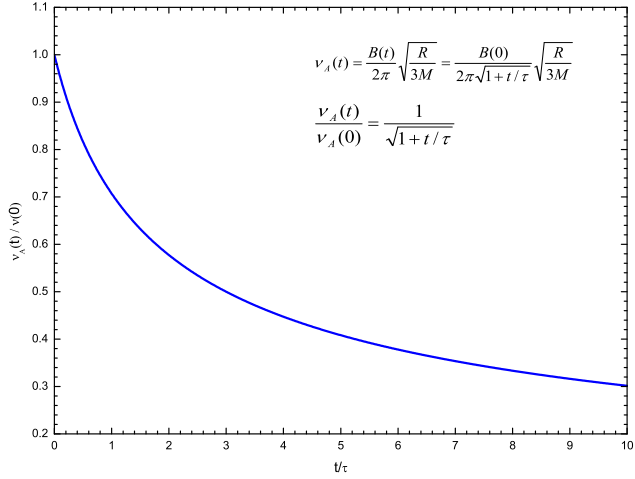}
  \caption{Time evolution of Alfv\'en frequency
  of Lorentz-force-driven vibrations converting vibration energy into power of magneto-dipole radiation.}\label{BPNS}
\end{figure}
 The difference between periods evaluated at successive moments of time $t_1=0$
 and $t_2=t$ is given by
 \begin{eqnarray}
 \label{e3.15}
 &&\Delta P(t)=P(t)-P(0)\\
 \nonumber
 &&=-P(0)\left[1-\frac{B(0)}{B(t)}\right]>0,\quad B(t)<B(0).
 \end{eqnarray}
 The practical usefulness of these general relations is that they can be
 used as a guide in search for
 fingerprints of Alfv\'en seismic vibrations in data on oscillating emission from
 quaking neutron star.

 \subsection{Time evolution of the vibration amplitude}

 The considered model permits
 exact analytic solution of equation for vibration amplitude $\alpha(t)$ which is convenient to represent as
\begin{eqnarray}
\label{ee1}
 &&{\ddot \alpha}(t)+\omega^2(t)\alpha(t)=0,\\
 \label{ee1a}
 && \omega^2(t)=\frac{\omega^2(0)}{1+t/\tau},\,\,\,\omega(0)=\omega_A\kappa.
  \end{eqnarray}
The procedure is as follows. Let us introduce new
variable $s=1+t/\tau$.
In terms of $\alpha(s)$, equation ({\ref{ee1}) takes the form
\begin{eqnarray}
\label{ee8}
 s\alpha''(s)+\beta^2\alpha(s)=0,\quad \beta^2=\omega^2(0)\tau^2={\rm const}.
  \end{eqnarray}
   This equation permits exact analytic solution\cite{PZ-04}
\begin{eqnarray}
\label{ee10}
 &&\alpha(s)=s^{1/2}\{C_1\,J_1(2\beta s^{1/2})+C_2Y_1(2\beta s^{1/2})\}
  \end{eqnarray}
 where $J_1(2\beta s^{1/2})$ and $Y_1(2\beta s^{1/2})$ are Bessel functions\cite{AS-72}
\begin{eqnarray}
\label{ee10a}
 && J_1(z)=\frac{1}{\pi}\int_0^\pi \cos(z\sin \theta-\theta)d\theta,\\
 \label{ee10b}
 && Y_1(z)=\frac{1}{\pi}\int_0^\pi \sin(z\sin \theta-\theta)d\theta,\quad z=2\beta s^{1/2}.
  \end{eqnarray}
The arbitrary constants $C_1$ and $C_2$ can be eliminated from two conditions
\begin{eqnarray}
\label{ee12}
 &&\alpha(t=0)=\alpha_0,\quad \alpha(t=\tau)=0.
  \end{eqnarray}
where zero-point amplitude
\begin{eqnarray}
\label{ee12a}
 &&\alpha_0^2=\frac{2{\bar E_A(0)}}{M\omega^2(0)}=\frac{2{\bar E_A(0)}}{K(0)},\quad
 \omega^2(0)=\frac{K(0)}{M}
  \end{eqnarray}
is related to the average energy
${\bar E}_A(0)$ stored in torsional Alfv\'en vibrations at initial (before magnetic field decay) moment of time $t=0$. Before magnetic field decay, the star oscillates in
 the harmonic in time regime, that is, with amplitude $\alpha=\alpha_0\cos(\omega(0) t)$, so that  $<{\alpha^2}>=(1/2)\alpha_0^2$  and $<{\dot \alpha^2}>=(1/2)\omega^2(0)\alpha_0^2$. The average energy of
 such oscillations is given by equation ${\bar E}_A(0)=(1/2)M<{\dot \alpha^2}>+
(1/2)K(0)<{\alpha^2}>=(1/2)M\omega^2(0)\alpha_0^2=(1/2)K(0)\alpha_0^2$
which relates the energy stored in vibrations with vibration amplitude $\alpha_0$.
As a result, the general solution of (\ref{ee1}) can be represented in the form
 \begin{eqnarray}
\nonumber
 &&\alpha(t)=C\,[1+(t/\tau)]^{1/2}\\
 \label{ee11}
 &&\times \{J_1(2\beta\,[1+(t/\tau)]^{1/2})-\eta\,Y_1
 (2\beta\,[1+(t/\tau)]^{1/2})\},\\
 \label{ee11a}
 && \eta=\frac{J_1(z(\tau))}{Y_1(z(\tau))},\quad C=\alpha_0[J_1(z(0))-\eta\,Y_1(z(0)]^{-1}.
  \end{eqnarray}
The vibration period lengthening in the process of vibrations is illustrated
in Fig.6 and Fig.7, where we plot $\alpha(t)$, equation (\ref{ee11}), at different values of parameters $\beta$ and $\eta$ pointed out in the figures. Fig.7
shows that $\eta$ is the parameter regulating magnitude of vibration
amplitude, the larger $\eta$, the higher  amplitude. However this parameter
does not affect the rate of period lengthening. As it is clearly seen from Fig.6
both the elongation rate of vibration period and magnitude of vibration
amplitude are highly sensitive to parameter $\beta$.
\begin{figure}
\centering
 \includegraphics[scale=0.5]{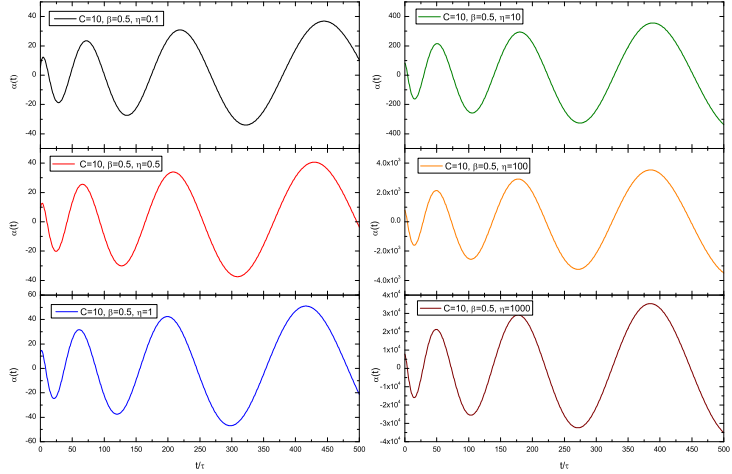}
 \caption{ Vibration amplitude $\alpha(t)$ computed at fixed $C$ and $\beta$
 and different values of $\eta$. This shows that variation of this parameter
 is manifested in change of magnitude of $|\alpha|$, but period elongation
 is not changed.}
 \end{figure}
 \begin{figure}
 \centering
 \includegraphics[scale=0.5]{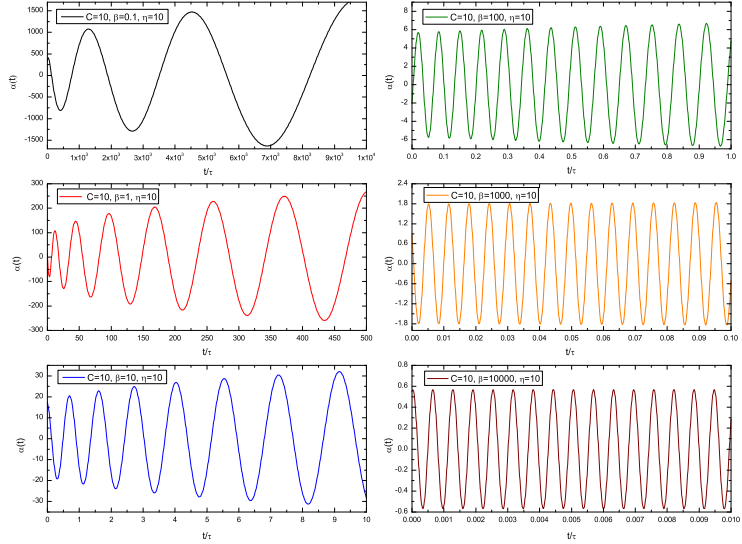}
 \caption{Vibration amplitude $\alpha(t)$ computed at fixed $C$ and $\eta$
 and different values of $\beta$.}
 \end{figure}

   All the above shows that the
   magnetic-field-decay induced loss of vibration energy is substantially
   different from the vibration energy dissipation caused by shear viscosity of matter
   resulting in heating of stellar material\cite{FS-92,M-99}.
   As was noted, the characteristic feature of this latter mechanism
   of vibration energy conversion into the heat (i.e., into the energy of non-coherent
   electromagnetic emission responsible for the formation of photosphere
   of the star)  is that the frequency and, hence, period of vibrations are the same
   as in the case of viscous-free vibrations\cite{B-10c}.
   However, it is no longer so in the case under consideration. It follows from
   above that depletion of magnetic field pressure resulting in the loss of total energy
   of Alfv\'en vibrations of the star causes its vibration period to lengthen at a rate
   proportional to the rate of magnetic field decay.

   \section{Numerical analysis}

To get an idea of ​​the magnitude ​​of characteristic parameter of vibrations providing 
energy supply of magneto-dipole radiation of a neutron star, in Table 1 we present results of numerical computations of the fundamental frequency of neutron star oscillations in quadrupole toroidal $a$-mode and time of decay of magnetic field $\tau$ as functions of increasing magnetic field.
As was emphasized, the most striking feature of considered model 
of vibration powered  radiation is the lengthening of periods of pulsating 
emission caused by decay of internal magnetic field. This suggests that this model 
 is relevant to electromagnetic activity of magnetars - neutron stars endowed
 with magnetic field of extremely high intensity the
 radiative activity of which is ultimately related to the magnetic field decay. 
 Such a view is substantiated by estimates of Alfv\'en frequency 
 presented in the table.   For magnetic fields of 
  typical rotation powered radio pulsars, $B\sim 10^{12}$ G, the computed frequency 
  $\nu_A$ is much smaller than the detected frequency of pulses whose origin is 
  attributed to lighthouse effect.  
  In the meantime, for neutron stars with magnetic fields 
  $B\sim10^{14}$ G the estimates of $\nu_A$  are in the realm of 
  observed frequencies of high-energy pulsating emission of soft gamma repeaters 
  (SGRs), anomalous X-ray pulsars (AXPs) and sources exhibiting similar 
  features. According to common belief, these are magnetars - highly magnetized 
  neutron stars whose radiative activity is related with magnetic field decay. The amplitude of vibrations is estimated
  as  \cite{RAA-11}
\begin{eqnarray}
\label{e3.1a}
&& \alpha_{0}= \left[\frac{2\bar E_{\rm A}(0)}{{\cal M}\omega^{2}(0)}\right]^{1/2}=3.423 \times 10^{-3} \bar E_{\rm A, 40}^{1/2} B^{-1}_{14} R_{6}^{-3/2}.
\end{eqnarray}
where $\bar E_{\rm A, 40}= \bar E_{\rm A}/(10^{40}\; {\rm erg})$ is the energy stored in the vibrations. $R_{6} = R/(10^{6} \; {\rm  cm})$ and $B_{14} = B/(10^{14} \; {\rm G})$. 
The presented computations show that the decay time of magnetic 
  field (equal to duration time of vibration powered radiation in question) strongly 
  depends on the intensity of initial magnetic field of the star: the larger 
  magnetic field $B$ the shorter time of radiation $\tau$ at the expense of energy 
  of vibration in decay during this time magnetic field. The effect of equation of state
  of neutron star matter (which is most strongly manifested in different values mass and radius of the star) on frequency $\nu_A$ is demonstrated by numerical vales of this quantity for magnetars with one and the same value of magnetic field $B=10^{15}$ G but different values of mass and radius.
  
   \begin{table}
\begin{center}
\caption{The Alfv\'en frequency of Lorentz-force-driven torsion vibrations, $\nu_A$, and their lifetime equal to decay time of magnetic field, $\tau$, in neutron stars with magnetic fields typical for pulsars and magnetars.}\label{TVT}
  \begin{tabular}{cccccc}
    \hline\hline
    & $M$($M_{\odot}$)  &$ R$(km)  & $B$(G)&  $\nu_{\rm A}$(Hz)  & $\tau$(yr)  \\
    \hline
  Pulsars &0.8 &20 &$10^{12}$ & $3.25\times 10^{-3}$ & $4.53 \times 10^{10}$\\
   & 1.0 & 15 & $10^{13}$ & $2.52 \times 10^{-2}$ &$2.98\times 10^{7}$  \\
  Magnetars &1.1 & 13& $10^{14}$ &0.22 & $7.4 \times 10^{3}$ \\
  &1.2 & 12& $10^{15}$ &2.06  &1.31  \\
  &1.3 & 11& $10^{15}$ &1.89  &2.38  \\
  &1.4 & 10 & $10^{16}$  &17.4  &$4.44 \times 10^{-4}$ \\ 
     \hline
  \end{tabular}\\
\end{center}
\end{table}
 
The 
luminosity powered by neutron star vibrations in quadrupole toroidal $a$-mode
is given by
\begin{eqnarray}
\label{e3.1b}
&& {\cal P}=\frac{\mu^2}{c^3} B^4(t) \alpha^2(t),\quad \mu=(1/2)B(0)R^3,\quad B(t)=B(0)[1+t/\tau]^{-1/2},\\
&& \alpha(t)=C\,s^{1/2}\{J_1(z(t))-\eta\,Y_1 (z(t))\},\quad z=2\omega(0)\tau\,(1+t/\tau)
\end{eqnarray} 
The presented in Fig.8 computations of power of magneto-dipole radiation 
of a neutron stars (of one and the same mass and radius but different values of magnetic fields) exhibit oscillating character of luminosity. 
The frequency of these oscillations equal to that of torsional Alfv\'en seismic vibrations of neutron star.

\begin{figure}
 \centering
 \includegraphics[scale=0.6]{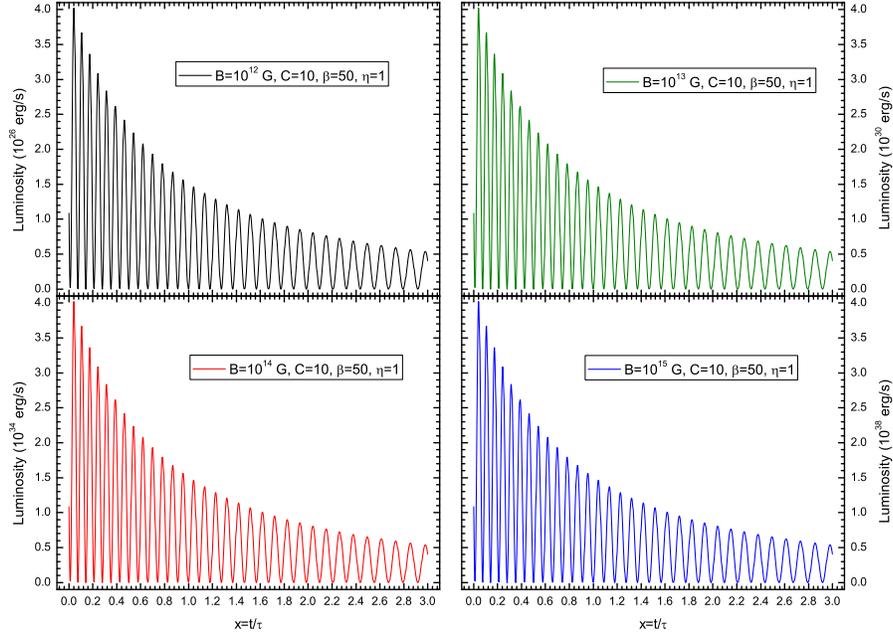}
  \caption{Time evolution of luminosity of magneto-dipole radiation 
  powered by energy of torsional Alfv\'en seismic vibrations 
  of neutron star with mass $M=1.2M_\odot$ and radius $R=15$ km 
  with intensities of magnetic field.}
  \label{fig:fr-period}
\end{figure}
  
  All above suggests that developed theory of vibration-energy powered emission of neutron star
 is relevant to electromagnetic activity of magnetars - neutron stars endowed
 with magnetic field of extremely high intensity the
 radiative activity of which is ultimately related to the magnetic field decay.
 This subclass of highly magnetized compact objects is commonly associated with soft gamma repeaters and anomalous X-ray pulsars\cite{H-99,K-99,WT-06,M-08}
  --  young  isolated and seismically active
 neutron stars\cite{C-96,F-00}. The magnetar quakes
 are exhibited by short-duration thermonuclear
 gamma-ray flash followed by rapidly oscillating X-ray flare of
 several-hundred-seconds duration. During this latter stage of quake-induced
 radiation  of magnetar, a long-periodic
 (2-12 sec) modulation of brightness was observed lasting for about 3-4 minutes. Such long period of pulsed emission
 might be expected from old rotation-energy
 powered pulsars (due to slow down of rotation), but not from magnetars which are young neutron stars, as follows
 from their association with pretty young supernovae. Taking this into account
 (and also the fact that energy release during X-ray flare is much larger than the
 energy of rigid-body rotation with so long periods) it has
 been suggested\cite{B-WS}  that detected long-periodic pulsating
  emission of magnetars is powered by  the energy of torsional magneto-elastic
  vibrations triggered by quake.  In view of key role of ultra-strong
  magnetic field it is
  quite likely that quasi-periodic oscillations (QPOs) of outburst flux from SGR
  1806-20 and SGR 1900+14 detected in\cite{I-05,WS-06,T-06} are produced
  by torsional seismic vibrations predominately sustained by Lorentz
  force\cite{B-09a,B-09b,B-10c} which are accompanied, as was argued
  above, by monotonic decay of background magnetic field. If so, the predicted
  elongation of QPOs period  of oscillating outburst emission from quaking
  magnetars should be traced in existing and future observations.

     \subsection{Comparison with the rotation powered neutron star}

For a sake of comparison, 
in the considered model of vibration powered radiation, the equation
of magnetic field evolution is obtained in similar fashion
  as that for the angular velocity $\Omega(t)$ does in
  the standard model of rotation powered neutron star which rests on equations \cite{ApSS-11}
  \begin{eqnarray}
  && \frac{dE_R}{dt}=-\frac{2}{3c^3}\delta {\ddot {\mbox{\boldmath
  $\mu$}}}^2(t),\\
  && E_R(t)=\frac{1}{2}I\,{\Omega}^2(t), \quad I=\frac{2}{5}MR^2
   \end{eqnarray}
  One of the basic postulates of the model of rotation-energy powered emission 
is that the time evolution of total magnetic moment of the star 
is governed by the equation 
  \begin{eqnarray}
  && \delta {\ddot {\mbox{\boldmath
  $\mu$}}}(t)=[\mbox{\boldmath ${\Omega}$}(t)\times [\mbox{\boldmath ${\Omega}$}(t)\times {\mbox{\boldmath
  $\mu$}}]],\quad {\mbox{\boldmath
  $\mu$}}={\rm constant}.
\end{eqnarray}
It follows 
\begin{eqnarray}
\label{ei2.4}
  && \delta {\ddot{\mbox{\boldmath ${\mu}$}}}^2(t)=\mu_{\perp}^2\Omega^4(t),\quad \mu_{\perp}=\mu\,\sin\theta
  \end{eqnarray}
 where $\theta$ is angle of inclination of $\mbox{\boldmath ${\mu}$}$ to $\mbox{\boldmath ${\Omega}$}(t)$. 
The total magnetic moment of non-rotating neutron star is parametrized by 
equation of uniformly magnetized (along the polar axis) sphere 
 \begin{eqnarray}  
{\mbox{\boldmath
  $\mu$}}=\mu {\bf n},\, \mu=\frac{1}{2}BR^3={\rm constant}.
 \end{eqnarray} 
 This parametrization presumes that in the rotation powered neutron star, 
 the frozen-in the star magnetic field operates like a passive promoter of magneto-
 dipole radiation, that is, intensity of the internal magnetic field remain constant in 
 the process of radiation. 
 As a result, the equation of energy conversion from rotation to magnetic dipole 
 radiation is reduced to equation of slow down of rotation 
\begin{eqnarray}
  &&{\dot \Omega}(t)=-K\Omega^3(t),\quad K=\frac{2\mu_{\perp}^2}{3Ic^3},\\
\label{e2.6a}
&& \Omega(t)=\frac{\Omega(0)}{\sqrt{1+t/\tau}},\quad \tau^{-1}=2\,K\,\Omega^2(0).
 \end{eqnarray}  
  where $\theta$ is angle of inclination of $\mbox{\boldmath  $\mu$}$  to $\mbox{\boldmath  ${\Omega}$}(t)$. From the last equation it follows 
  \begin{eqnarray}
 && P(t)=P(0)\,[1+(t/\tau)]^{1/2},\,\,  P(0)=\frac{2\pi}{\Omega(0)},\\
 && {\dot P}(t)=\frac{1}{2\tau}\frac{P(0)}{[1+(t/\tau)]^{1/2}}
 \end{eqnarray}
  and, hence, the lifetime $\tau$ is related with $P(t)$ and ${\dot P}(t)$ 
  as \begin{eqnarray}
  \label{e2.6b}
 && {P(t)}{\dot P}(t)=\frac{P^2(0)}{2\tau}={\rm constant}.
 \end{eqnarray}
  Equating two independent estimates for $\tau$ given by  
  equations (\ref{e2.6a}) and  (\ref{e2.6b}) we arrive at 
  widely utilized analytic estimate of magnetic field on the neutron star pole:
  $$B=[3Ic^3/(2\pi^2\,R^6)]^{1/2}\sqrt{P(t)\,\dot P(t)}.$$
  For a neutron star 
  of mass $M=M_\odot$, and radius $R=13$ km, one has 
  $$B=3.2\,10^{19} 
  \sqrt{P(t)\,\dot P(t)},\,\,{\rm G}.$$
  The outlined treatment of rotation-powered magneto-dipole radiation of a neutron  
  star emphasizes kinematic nature of variation of magnetic moment of pulsar whose 
  magnetic field is regarded as independent of time.
 Thus, the substantial physical difference between vibration-powered and 
 rotation-powered neutron star models is that  in the former the elongation of pulse period is attributed to magnetic field decay, whereas in the latter the period lengthening 
  is ascribed to slow down of 
  rotation \cite{MT-77,LK-04,BK-10}. 

  \subsection{Comment on magnetic field decay in quaking neutron star}

    The considered law of magnetic field decay cannot be, of course, regarded as universal because it reflects a quite concrete line of argument regarding the
fluctuations of magnetic moment of the star. The interrelation between quake-induced oscillations of total magnetic moment  of the star,
$\delta \mbox{\boldmath $\mu$}(t)$, and the amplitude, ${\alpha}(t)$, of its seismic magneto-mechanical oscillations can be consistently interpreted
with the aid of the function of dipole demagnetization ${\bf f}(B(t))$ which is defined by the following condition of self-consistency in $\alpha$ of
 right and left hand sides of equation (\ref{e3.4}), namely
 \begin{eqnarray}
  \label{e2.18a}
  \delta {\ddot {\mbox{\boldmath $\mu$}}}(t)=\,{\bf f}(B(t)){\alpha}(t).
  \end{eqnarray}
 The vector-function of dipole demagnetization ${\bf f}(B(t))$  depending on decaying magnetic field reflects temporal changes of electromagnetic properties of
 neutron star matter as well as evolution of magnetic-field-promoted coupling between neutron star and its environment. This means that specific form
 of this phenomenological function should be motivated by heuristic arguments taking into account these factors. With this form of  $\delta {\ddot
 {\mbox{\boldmath $\mu$}}}(t)$, equation (\ref{e3.4}) is  transformed to magnetic field decay of the form
  \begin{eqnarray}
 \label{e2.19a}
  && \frac{{\rm d}B(t)}{{\rm d}t}=-\eta\,\frac{{\bf f}^2(B(t))}{B(t)},\quad
  \eta=\frac{2}{3{\cal M}\kappa^2c^3}={\rm const}.
  \end{eqnarray}
  In the above considered case, this function is given by
  \begin{eqnarray}
  \label{e2.20a}
  && {\bf f}(B(t))=\beta\,B(t)\,{\bf B}(t),\quad \beta= \kappa^2\mu={\rm
  constant}.
  \end{eqnarray}
The practical usefulness of the dipole demagnetization function, ${\bf f}(B(t))$, consists in that it provides economic way of studying a vast variety of
heuristically motivated laws of magnetic field decay, $B=B(t)$, whose inferences can ultimately be tested by observations.
      In this subsection,  with no 
discussing any specific physical mechanism which could be responsible for magnetic field decay, we consider a set of representative examples
of demagnetization function ${\bf f}(B(t))$  some of which have been regarded before, though in a somewhat different context [42-49].
Here we stress again that the model under consideration deals with magnetic field decay in the course of vibrations triggered by starquake, not with long-term secular decay
which has been the subject of these latter investigations (see also references therein).

 \begin{figure*}[ht]
\begin{center}
 \includegraphics[width=12cm]{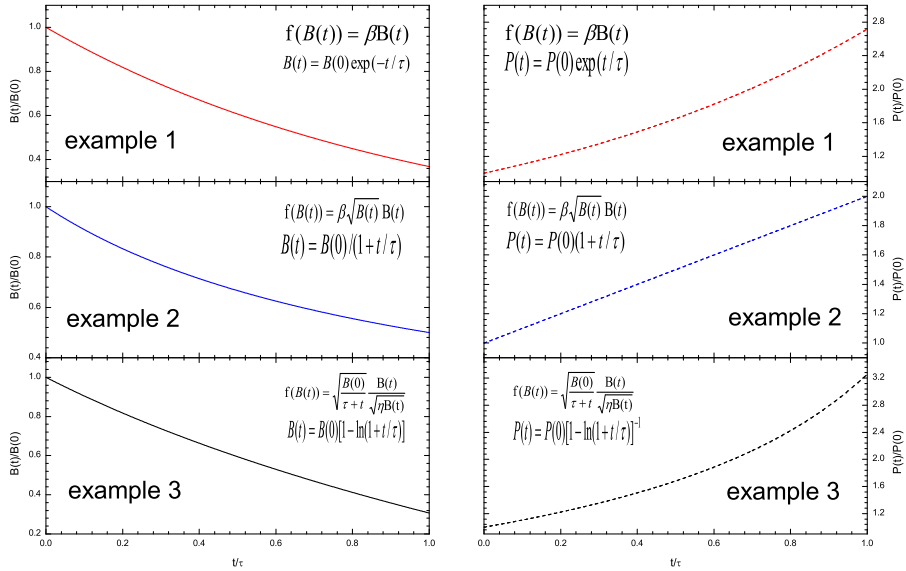}
 \end{center}
\caption{Three representative examples with different laws of the magnetic field decay resulting in the elongation of the vibration
period for each case.}\label{EP}
\end{figure*}

1. As a first representative example, a model of quaking neutron star whose function of dipole demagnetization has the following form
 \begin{eqnarray}
 \label{a.1}
 {\bf f}(B(t))=\beta {\bf B}(t)
 \end{eqnarray}
is been considered. In such demagnetization model, the temporal evolution of the magnetic field and the
lengthening of the vibration period obey the following laws
 \begin{eqnarray}
 \label{a.2}
&&  B(t)=B(0)\,e^{-t/\tau},\,\,P(t)=P(0)\,e^{t/\tau},\,\,
\tau = \frac{P(t)}{\dot{P}(t)}.
 \end{eqnarray}
 where $B(0)$ is the intensity of magnetic field before quake.
  2.  For a quaking neutron star model whose function of demagnetization is given by
  \begin{eqnarray}
 \label{a.3}
 {\bf f}(B(t))=\beta\sqrt{B(t)}\, {\bf B}(t)
 \end{eqnarray}
the resultant equation of magnetic field evolution and the vibration period elongation read
 \begin{eqnarray}
 \nonumber
  &&B(t)=B(0)\left(1+\frac{t}{\tau}\right)^{-1},\\
   \label{a.4}
  && P(t)=P(0)\left(1+\frac{t}{\tau}\right),\,\, \tau = \frac{P(0)}{\dot{P}(0)}.
 \end{eqnarray}
 Similar analysis can be performed
 for the demagnetization function of the form
  \begin{eqnarray}
 \nonumber
 {\bf f}(B(t))=\beta\sqrt{B^{m-1}(t)}\, {\bf B}(t), \quad m =3,4,5...
 \end{eqnarray}
 Namely,
 \begin{eqnarray}
 \nonumber
 &&\frac{dB(t)}{dt}=-\gamma\,B^{m}(t),\quad \gamma=\eta\beta^2,\\
 \nonumber
 && B(t)=\frac{B(0)}{[1+t/\tau_m]^{1/(m-1)}},\quad \tau_m^{-1}=\gamma
 (m-1)B^{m-1}(0).
 \end{eqnarray}

 3. Finally, let's consider a model with quite sophisticated function of dipole demagnetization
 \begin{eqnarray}
 \label{a.5}
 && {\bf f}(B(t))=\sqrt{\frac{B(0)}{\tau+t}}\frac{{\bf B}(t)}{\sqrt{\eta B(t)}}.
  \end{eqnarray}
 which lead to a fairly non-trivial logarithmic law of magnetic field decay and vibration period lengthening
  \begin{eqnarray}
 \label{a.6}
  &&  B(t)=B(0)\left[1-\ln\left(1+\frac{t}{\tau}\right)\right],\\
   \label{a.6a}
  && P(t)=P(0)\left[1-\ln\left(1+\frac{t}{\tau}\right)\right]^{-1},\quad
   \tau =\frac{P(0)}{{\dot P}(0)}.
 \end{eqnarray}
 For each of above examples, the  magnetic field and the resultant lengthening of vibration period, computed as  functions of fractional time $t/\tau$, are shown in Fig. 9.
This third  example is interesting in that computed in this model
ratio ${\dot \nu}(t)/\nu(0)$, pictured in Fig.10, as a function of $t$, is similar to that which exhibit data on post-glitch emission of PSR J1846-0258 [50].
 \begin{figure}
 \centering
 \includegraphics[scale=0.5]{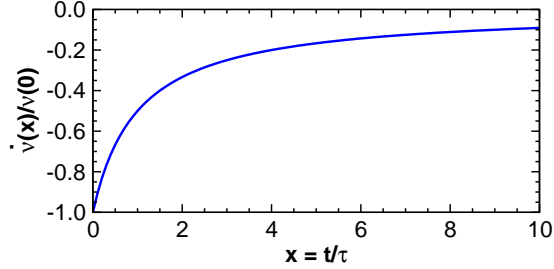}
  \caption{
 The rate of frequency  ${\dot \nu}(x)$ normalized to $\nu(0)$ as a function of
 $x=t/\tau$ for logarithmic law of magnetic filed decay.}
 \end{figure}
These examples show that the period elongation is the common effect of  magnetic field decay. The interrelations between periods and its
derivatives substantially depend on specific form of the magnetic field decay law although physical processes responsible for magnetic field decay remain uncertain.
Practical significance of above heuristic line of argument is that it leads to meaningful conclusion (regarding elongation of periods of oscillating magneto-dipole emission)
even when detailed mechanisms of magnetic field decay
in the course of vibrations are not exactly known.

\section{Summary}

It is generally realized today that the standard model of 
 inclined rotator, lying at the base of our understanding of radio pulsars, faces 
 serious difficulties in explaining  the long-periodic ($2<P<12$ s) pulsed radiation of 
 soft gamma repeaters (SGRs) and anamalous $X$-ray pulsars (AXPs). 
 Observations show that persistent $X$-ray luminosity of these sources      
 ($10^{34}<L_X<10^{36}$ erg  s$^{-1}$) is  appreciably (10-100 times) larger 
 than expected from neutron star deriving radiation power from energy of rotation 
 with frequency of detected pulses. It is believed that this discrepancy can be resolved 
 assuming that AXP/SGR-like sources 
 are magnetars -- young, isolated and seismically active neutron stars whose energy supply of pulsating high-energy radiation comes not from 
 rotation (as is the case of radio pulsars) but from different process
 involving decay of ultra strong magnetic field,  $10^{14}<B<10^{16}$ G.
 Adhering to this attitude we have presented the model of quaking  neutron star 
 deriving radiation power from the energy of torsional Lorentz-force-driven oscillations. It is
 appropriate to remind early works of the infancy of neutron star era  
 \cite{HNW-64,C-65}
 in which it has been pointed out for the first time that vibrating neutron star should
 operate like Hertzian magnetic dipole deriving radiative power of magneto-dipole
 emission from the energy of magneto-mechanical vibrations\cite{P-08}.
 What is newly disclosed 
 here is that the main prerequisite of the energy conversion from vibrations into radiation is the decay of magnetic field in the star.
 Since the magnetic field decay is one of the most conspicuous 
 features distinguishing magnetars from rotation powered 
 pulsars,  it seems meaningful to expect that at least some of AXP/SGR - like sources  
 are magnetars deriving power of pulsating magnetic dipole radiation from the energy 
 of quake-induced torsion seismic vibrations.

\section*{Acknowledgment}

    This work is supported by the National Natural Science Foundation of China
(Grant Nos. 10935001, 10973002), the National Basic Research Program of
China (Grant No. 2009CB824800), the John Templeton Foundation and also 
National Science Council of Taiwan, grant number NSC 99-2112-M-007-017-MY3.

\end{document}